\documentclass[12pt]{article}
\usepackage{ifpdf}
\pdfoutput=1
\usepackage[a4paper,width=18cm, bottom=3.2cm, left=1.5cm]{geometry}
\usepackage{fancyhdr}
\usepackage{amssymb}
\usepackage{amsmath}
\usepackage[utf8]{inputenc}
\usepackage[polish,english]{babel}
\usepackage{polski}
\usepackage{graphicx,caption}
\usepackage{hyperref}
\usepackage{color}

\usepackage{microtype}

\usepackage{authblk}

\begin{document}

\def\lsim{\stackrel{\scriptstyle <}{\phantom{}_{\sim}}}
\def\gsim{\stackrel{\scriptstyle >}{\phantom{}_{\sim}}}

\title{\bf  Formation of hadrons at chemical freeze-out}

\author{David Blaschke
\footnote{Email:david.blaschke@ift.uni.wroc.pl}}
\author{Jakub Jankowski
\footnote{Email:jakubj@th.if.uj.edu.pl}}
\author{Michał Naskręt
\footnote{Email:michal.naskret@uwr.edu.pl }}
\affil{Institute of Theoretical Physics, University of Wroclaw, pl. Maxa Borna 9, 50-204  Wroc{\l}aw, Poland}

\linespread{1.4}
\parskip=6pt
\parindent=0pt

\maketitle

\thispagestyle{empty}

\abstract{ 
We use a kinetic condition to predict the chemical freeze-out 
parameters for hadronic species produced in heavy ion collisions. 
The resulting freeze-out lines 
for different hadrons lie close to one another in the temperature and
baryochemical potential plane, defining a universal, narrow region.
The chemical freeze-out is driven by the localization of hadrons
due to the chiral symmetry breaking and confining aspects of
the QCD transition.
}

\newpage


\section{Introduction}

The systematic investigation of particle production in heavy-ion collisions (HIC) over the past thirty years
has brought a number of surprising discoveries. 
One of them is the simple behaviour of produced particle yields captured within the elementary statistical model \cite{Adamczyk:2017iwn,BraunMunzinger:2001ip,Broniowski:2001we,Petran:2013lja,Petran:2013qla,Cleymans:2005xv,Andronic:2005yp}, see Ref.~\cite{BraunMunzinger:2003zd} for an introduction. 
Particle abundances are described by the equilibrium distribution functions and determined by three parameters: the temperature $T_f$, the baryochemical potential $\mu_f$ and the system's volume.
This effect is referred to as the chemical {\it freeze-out}.
In a wide range of energies the freeze-out parameters are extracted from the experimental data, and provide a landmark in the QCD phase diagram. 
Their relation to the QCD  transition remains to be uncovered. 
A universal fit gives the dependence of the $T_f$ and $\mu_f$ parameters on the centre of mass energy
of colliding nucleons  \cite{Cleymans:2005xv,Andronic:2005yp}. 
In addition, extensive studies reveal a number of phenomenological relationships obeyed at the 
freeze-out \cite{Cleymans:2005xv,Petran:2013qla}. 
	 
Despite the simple behaviour it is very hard to uncover the underlying physical mechanism of the freeze-out phenomenon and its relations to the QCD transition.  
A quantitative determination of the freeze-out parameters can be realised by comparing the ratios of net electric charge fluctuations computed within lQCD \cite{Bazavov:2012vg} with the measured net proton number fluctuations \cite{Aggarwal:2010wy}.
While being based on first principle computations this procedure does not reveal the detailed mechanism of the hadronic freeze-out, which is crucial in the context of the search for the QCD critical point and  recent experimental results \cite{HADES}:
the low energy point determined in the HADES experiment is significantly off the parametric line determined by the fit to other available data \cite{Cleymans:2005xv}.
	
In this letter we attempt to construct a simple model for the freeze-out phenomenon which provides a direct relation between the hadronic multiplicities and the QCD transition. 
The main assumption is that the chemical composition of the expanding, strongly interacting matter is fixed in the transition to the confined phase due to rapid formation of hadrons. 
We estimate hadron sizes based on a geometric law for hadron-hadron cross section, which we describe below. 
The hadron formation is meant here as a shrinking of the transverse hadron radius to the value attained in the vacuum.
This process can be intuitively viewed as an incarnation of the Mott-Anderson localisation known from solid state physics \cite{MOTT:1968zz}.
In a way it is the reverse process to the Mott dissociation of hadrons that occurs in the stage of compressing and heating nuclear matter in a heavy ion collision \cite{Satz:1983jp,Blaschke:1984yj}. 
However, we have to stress that in the present contribution we do not have a non-trivial model for the hadron wave functions, and we only use the geometric scaling law for the cross sections (see below).
It turns out that our model predictions tend to grasp the newly measured HADES point, along with the points extracted from high energy runs.
	
To make quantitative predictions we propose a simple kinetic freeze-out condition of equating the time scale related to the system's expansion rate with that of the flavour changing collisions \cite{Blaschke:2011ry,Blaschke:2011hm} leading to chemical equilibration.
This is very similar in spirit to the Knudsen number condition for the breakdown of the hydrodynamic description of time evolution of matter produced in HIC \cite{Niemi:2014wta,Ahmad:2016ods}, which has been recently used in the context of the dynamical freeze-out determination providing predictions 
for RHIC and LHC energies \cite{Ahmad:2016ods}. 
In this contribution we adopt a simpler approach, which allows us to address a wider energy range, at the prize of less detailed knowledge of particle dynamics.
We estimate the expansion rate from an entropy conserving flow. 
For the collision time we adopt a formula based on the geometric Povh-H{\"u}fner law for hadron-hadron cross sections \cite{Povh:1990ad,Hufner:1992cu}.
On top of that, we assume that hadron radii inherit a medium dependence from that of the chiral condensate as motivated by a Nambu--Jona-Lasinio (NJL) model calculation \cite{Hippe:1995hu} and the 
Gell-Mann-Oakes-Renner (GMOR) relation \cite{GellMann:1968rz}. 
The deconfining transition has its impact on the temperature dependence of the string tension determining the cross section. 
Within a range of model parameters we obtain a quantitative description of the freeze-out line in the temperature-baryochemical potential plane \cite{Cleymans:2005xv}.
Despite the limitations our model provides for the first time a direct relation of the freeze-out with the main effects accompanying the QCD transition.
	
The paper is organized as follows. 
In section \ref{model} we carefully expose and discuss all model assumptions.
Section \ref{results} contains a discussion of our predictions in the light of known experimental and phenomenological knowledge. 
Section \ref{conclusions} contains a brief summary and perspective for future research. 


\section{The freeze-out}
\label{model}

Our main model assumption is that freeze-out takes place when the expansion rate of the fireball is equal to the time scale of flavour-changing hadron-hadron collisions, i.e., 
\begin{equation}
H_{\text{exp}}(T,\mu)=\tau_{\text{coll},i}^{-1}(T,\mu)~,
\label{eq:Hexp}
\end{equation}
where $T$ is the temperature and $\mu$ the baryochemical potential of the hadron gas. 
Here, $H_{\text{exp}}(T,\mu)$ is the expansion rate, an analogue of the Hubble parameter
known from cosmology as
\begin{equation}
H_{\text{exp}}(\tau)= \frac{\dot{R}(\tau)}{R(\tau)}~,
\label{eq:}
\end{equation}
where $R(\tau)$ is the scale factor of the system at the time $\tau$ in its evolution,
and in turn the system's volume is $V(\tau)\sim R(\tau)^3$.
The trajectory $\tau(T,\mu)$ of the evolution in the phase space is obtained assuming entropy conserving flow 
$S=s(T,\mu)V(\tau_{\text{exp}})={\rm const}$,
and simple volume scaling $V(\tau_{\text{exp}})\propto \tau_{\text{exp}}^3$ resulting in
\begin{equation}
H^{-1}_{\rm exp}(T,\mu)=\tau_{\text{exp}}(T,\mu)=as^{-1/3}(T,\mu)~,
\label{eq:}
\end{equation}
where $\tau_{\text{exp}}(T,\mu)$ is total expansion time, and $a$ is a dimensionless constant of the order $1$ and is a measure if the initial state entropy. 
We further assume that the entropy density is that of a free hadron gas with contributions from individual hadron species $i$ given by
\begin{eqnarray}
  s^{B}_i(T,\mu_i)&=&\xi_{B}\frac{d_i}{2\pi^2}\int_{0}^{\infty}p^2dp\left(\ln(1-\exp[-(E_i-\mu_i)/T])-\frac{E_i-\mu_i}{T(\exp[(E_i-\mu_i)/T]- 1)}\right)\, ,
\\
s^{F}_i(T,\mu_i)&=&\xi_{F}\frac{d_i}{2\pi^2}\int_{0}^{\infty}p^2dp\Bigg(\ln(1+\exp[-(E_i-\mu_i)/T])+\frac{E_i-\mu_i}{T(\exp[(E_i-\mu_i)/T]+1)}
	\nonumber\\
  &&+\ln(1+\exp[-(E_i+\mu_i)/T])+\frac{E_i+\mu_i}{T(\exp[(E_i+\mu_i)/T]+1)}\Bigg)~.
\label{eq:}
\end{eqnarray}
The total entropy density is the sum of all individual terms
$s=\sum_i s_i$,
where $E_i=\sqrt{p^2+m_i^2}$, $\mu_i=\mu B_i+\mu_SS_i$, $a=2.86$, $\xi_{F}=4.2$ and $\xi_{B}=3.6$.

On the other hand we have to specify the rates of the relevant hadronic reactions. 
Those, in turn, determine the collision time scale relevant for the assumed freeze-out 
condition Eq.~(\ref{eq:Hexp}). 
For an individual hadron $i$ we have
\begin{equation}
\tau^{-1}_{\text{coll},i}(T,\mu)=\sum_{j}\sigma_{ij} v_{\rm rel} n_j(T,\mu)~,
\label{eq:}
\end{equation}
%
where $v_{\rm rel}$ is the relative velocity fixed to be $v_{\rm rel}=1$.
In the above $n_i(T,\mu)$ is a scalar density given by
\begin{equation}
n_{i}(T,\mu)=\frac{d_i}{2\pi^2}\int_{0}^{\infty}dp \, p^2\frac{m_i}{E_i}\frac{1}{\exp(E_i/T) - 1}~,
\label{eq:}
\end{equation}
for mesons and
\begin{eqnarray}
n_{i}(T,\mu)=\frac{d_i}{2\pi^2}\int_{0}^{\infty}dp \, p^2\frac{m_i}{E_i}\left(\frac{1}{\exp(E_i/T - B\mu/T - S\mu_{\text{S}}/T)+ 1}\right.\\
\nonumber
+\left.\frac{1}{\exp(E_i/T + B\mu/T + S\mu_{\text{S}}/T) + 1}\right)~,
\label{eq:scalarFermion}
\end{eqnarray}
for baryons. 
Close to the deconfining transition the most dominant reactions are the inelastic, flavour changing
processes. 
Therefore we can use the approximate relation
$\sigma_{\rm in-el}\simeq\sigma_{\rm tot}$. 
The elastic part becomes relevant for lower temperatures and/or densities.
The total hadron-hadron cross sections are modelled by the Povh-H{\"u}fner law (also referred to as a geometric scaling law) \cite{Povh:1990ad,Hufner:1992cu}
\begin{equation}
\sigma_{ij}(T,\mu)=\lambda\langle r_i^2\rangle_{T,\mu} \langle r_j^2 \rangle_{T,\mu}~,
\label{eq:CrossSection}
\end{equation}
where $\langle r_i^2\rangle_{T,\mu}$ is the medium dependent, mean squared radius of a hadron of species $i$, and $\lambda$ is the medium dependent string tension. 
This law has been established for the hadron-hadron reactions at zero temperature and zero density. 
Here we assume its validity for the in-medium hadronic reactions, attributing the medium dependence 
to all quantities appearing in Eq.~(\ref{eq:CrossSection}) (see below).
The original vacuum data were analysed for high energy collisions, i.e., $\sqrt{s}\gtrsim 15$~GeV, with 
the vacuum string tension value $\lambda_0=0.197$ GeV$^{-2}$ \cite{Hufner:1992cu}. 
Interestingly, non-relativistic potential models for the quark exchange contributions to hadron-hadron interactions suggest that this geometric law holds also for lower energies 
\cite{Martins:1994hd,Barnes:1991em}. 
The cross section (\ref{eq:CrossSection}) has a vacuum limit which corresponds to the measured hadronic reaction rates \cite{Povh:1990ad,Hufner:1992cu}.

In the next step we need to specify how hadron radii depend on the medium variables.
We do that by specifying the radii for pions, kaons, nucleons and lambda baryons and then using the 
same formulas for other hadrons according to their quark flavour content. 
This approach clearly looses dynamical information related for example with spin, as for example 
$\rho$ and $\omega$ mesons will have the same cross sections as the pion.

\begin{figure}[h]
	\begin{center}
	\includegraphics[height=.41\textheight]{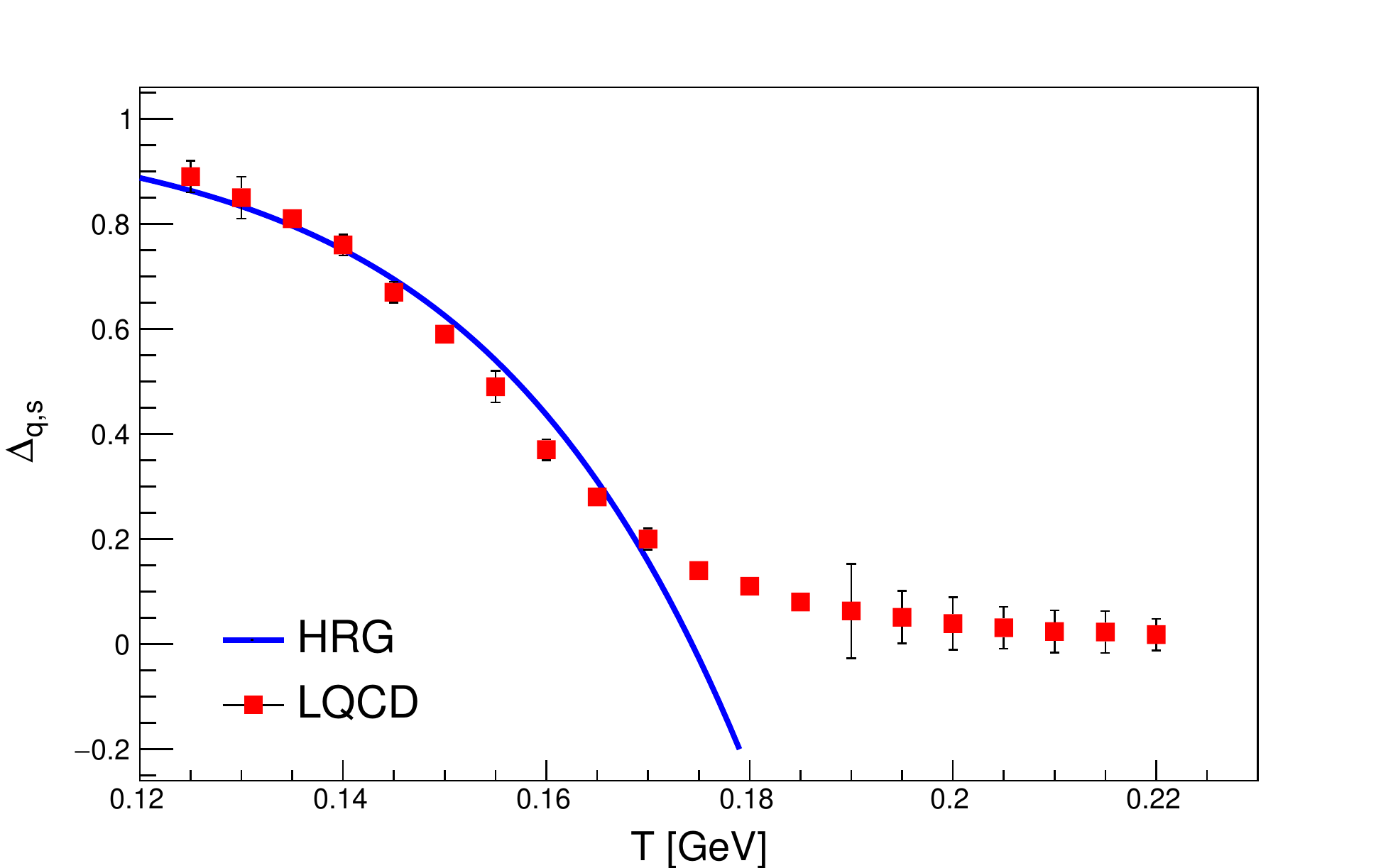}
	\caption{The temperature dependence of the chiral condensate $\Delta_{q,s}$, Eq. (\ref{eq:Delta})  at zero chemical potential
	for $(2+1)$-flavour  lattice QCD (squares) \cite{Borsanyi:2010bp} and
 	the hadron resonance gas model supplemented with NJL dynamics (solid line) \cite{Jankowski:2012ms}.} 
	\label{fig:Condensate}
	\end{center}
\end{figure}

For the pion radius the medium dependence  is motivated by the NJL-model prediction 
\cite{Hippe:1995hu} for the part that dominates in the vicinity of the chiral restoration transition
\begin{equation}
\langle r^2_{\pi} \rangle_{T,\mu}=\frac{3}{4\pi^2 f_{\pi}^2}~.
\label{eq:rpi1}
\end{equation} 
For the nucleon we adopt an effective radius composed of a core radius $r_0=0.45$ fm and 
the radius of the pion cloud from (\ref{eq:rpi1})
\begin{equation}
\langle r^2_{\rm N} \rangle_{T,\mu} = r_0^2 + \langle r_{\pi}^2\rangle_{T,\mu}~.
\label{eq:}
\end{equation}
Assuming that the Gell-Mann-Oakes-Renner relation holds in the hadronic phase also at finite 
temperature and density 
\begin{equation}
f_{\pi}^2(T,\mu) = \frac{-m_q\langle \bar q q \rangle_{T,\mu}}{m_{\pi}^2}~,
\label{eq:GMORpi}
\end{equation}
we can trade the pion decay constant $f_\pi$ for the chiral condensate
\begin{equation}
\langle r^2_{\pi}\rangle_{T,\mu}=\frac{3m_{\pi}^2}{4\pi^2 m_q}|\langle \bar q q \rangle_{T,\mu}|^{-1}~.
\label{eq:}
\end{equation}
For the kaon the corresponding formulas read
\begin{equation}
f_K^2m_K^2 = - \frac{\langle\bar{q}q\rangle_{T,\mu} + \langle\bar{s}s\rangle_{T,\mu}}{2}(m_q+m_s)~.
\label{eq:GMORkaon}
\end{equation}
\begin{equation}
\langle r^2_{\rm K} \rangle_{T,\mu} =\frac{3}{4\pi^2 f_{\rm K}^2} 
=
\frac{3}{2\pi^2}\frac{m_{\rm K}^2}{|\langle\bar{q}q\rangle_{T,\mu} + \langle\bar{s}s\rangle_{T,\mu}|(m_q+m_s)} ~.
\label{eq:rpi1}
\end{equation} 
For strange baryons, like the $\Lambda$ hyperon, we define the corresponding radius
\begin{equation}
\langle r_\Lambda^2\rangle_{T,\mu} = r_0^2 + \langle r_{\rm K}^2\rangle_{T,\mu}~.
\label{eq:}
\end{equation}
In this picture hadron radii are determined by the medium dependence of the quark condensates. 
To quantify their temperature and density dependence we use the hadron resonance gas model (HRG) supplemented with hadronic sigma terms computed in the NJL model \cite{Jankowski:2012ms}.
For a comparison to lattice QCD data we use the quantity \cite{Borsanyi:2010bp}
\begin{equation}
\Delta_{\rm q, s}(T,\mu) = \frac{\langle \bar{q}q\rangle - \frac{m_q}{m_s}\langle \bar{s}s\rangle}{\langle \bar{q}q\rangle_0 - \frac{m_q}{m_s}\langle \bar{s}s\rangle_0}~.
\label{eq:Delta}
\end{equation}
The parameters are chosen in the following way:
$m_q=5.5$ MeV, $m_s=138$ MeV, $\langle\bar{q}q\rangle_0=(240~{\rm MeV})^3$
and $\langle\bar{s}s\rangle_0=0.8\langle\bar{q}q\rangle_0$ in 
accordance with lattice data \cite{Borsanyi:2010bp}.
The results for the temperature dependence of (\ref{eq:Delta}) are shown in Fig.~\ref{fig:Condensate}
and compared to the lattice data of Ref.~\cite{Borsanyi:2010bp}. 


\begin{figure}[h]
	\begin{center}
	\includegraphics[height=.41\textheight]{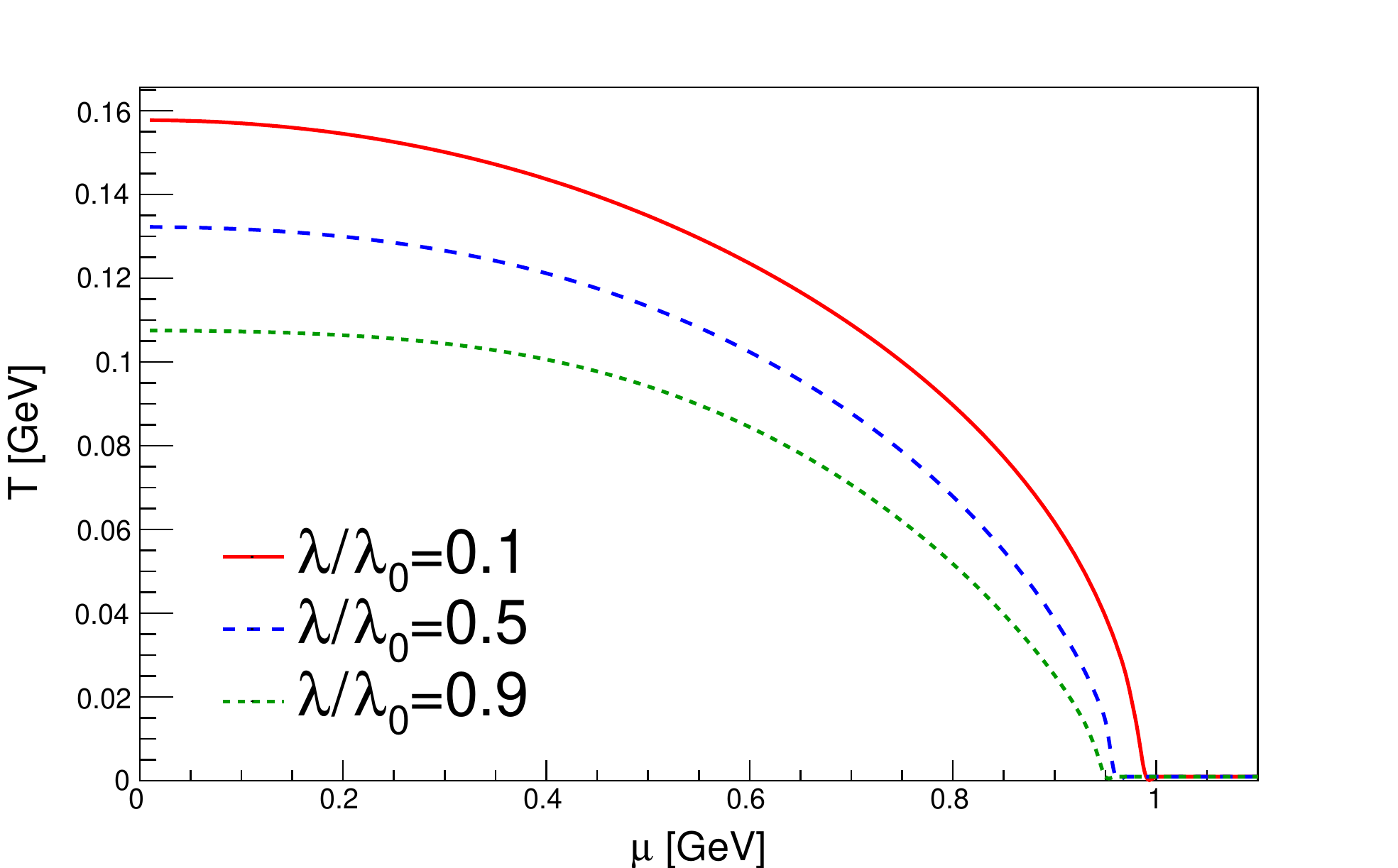}
	\caption{The temperature and chemical potential dependence of the string tension as defined 
	by Eq.~(\ref{eq:LambdaLorentz}). In the plot $\lambda_0=0.197$ GeV$^{-2}$ \cite{Hufner:1992cu}.}
	\label{fig:lambda}
	\end{center}
\end{figure}

Another effect driving the freeze-out in our model is colour confinement. 
We model the effect of approaching the deconfining transition by adopting for the string tension a phenomenological formula
\begin{equation}
\lambda(T,\mu)=\frac{\lambda_0}{1+\left(n_S(T,\mu)/n_0)\right)^\alpha}~,
\label{eq:LambdaLorentz}
\end{equation}
where $n_S=\sum_i n_i$ is the total scalar density, $\lambda_0=0.197$ GeV$^{-2}$ the vacuum value of the string tension, $\alpha\sim2$ and $n_0=0.08$ fm$^{-3}$.
This parametrization takes into account the string melting effects as a consequence of the deconfining transition of QCD.
In contrast to the chiral condensate there are no lattice data for the string tension for full QCD, which would include $(2+1)$ flavours of quarks. 
The only guidelines are quenched lattice results performed by the Bielefeld group \cite{Kaczmarek:1999mm,Doring:2007uh}.
The important aspect we expect to appear in full QCD, is that the onset of the "string melting" occurs 
for temperatures and densities where the string tension is $90\%$ of its vacuum value (lower dashed line 
in Fig.~\ref{fig:lambda}). 


\section{Results for the freeze-out conditions}
\label{results}


\begin{figure}[h]
	\begin{center}
	\includegraphics[height=.41\textheight]{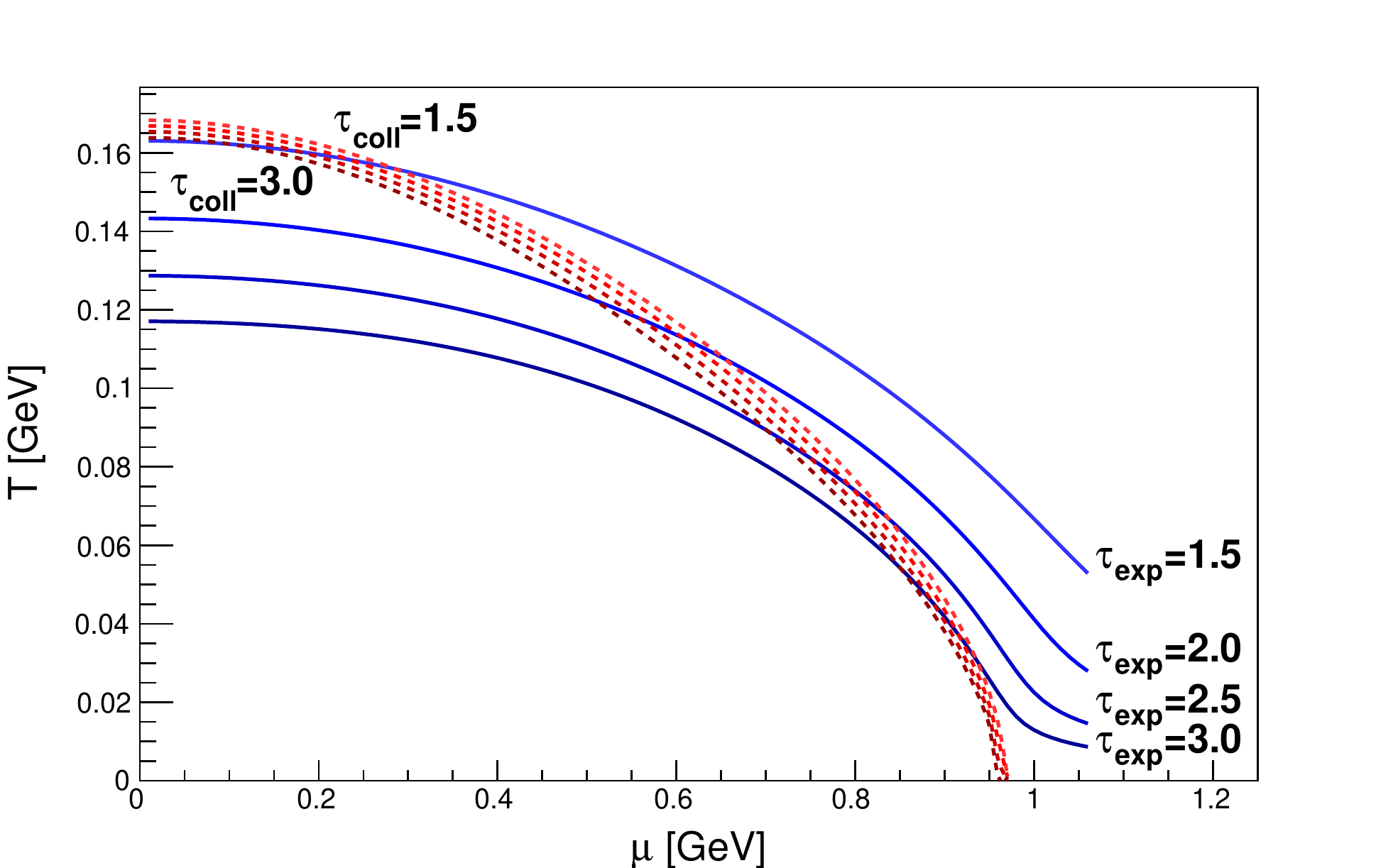}
	\caption{Expansion times $\tau_{\rm exp}=1.5(0.5)3.0~{\rm fm/c}$ (solid lines) and pion collision times
	 $\tau_{\rm coll}=1.5(0.5)3.0 ~{\rm fm/c}$  (dashed lines) 
	 as a function of temperature and baryochemical potential.}
	\label{fig:TimeScalesDiagram}
	\end{center}
\end{figure}


In this section we compare our model predictions for the freeze-out conditions computed for different hadron species with the values extracted from experiments.
As it was already explained, for every hadron species we define the corresponding freeze-out line based on Eq.~(\ref{eq:Hexp}). 
The most important prediction of our model is that the species dependent freeze-out lines lie close to one another, defining a narrow region in the $T-\mu$ plane. 
This is possible because the collision times for individual hadrons undergo a rapid change in a narrow region of temperatures and densities, which is a consequence of geometric localisation.
We illustrate this effect in Fig.~\ref{fig:TimeScalesDiagram}, where we show the $T$ and $\mu$ 
dependence of the expansion times along with the pion collision time.
While the former quantity has a rather smooth dependence, the latter one exhibits a sudden drop. 
This effect is triggered by the changes in the chiral condensate together with the modification of the string tension occurring in the range of temperatures and densities just around the QCD transition line. 
The intersection line of the two surfaces for collision time and expansion time in the $T-\mu$ plane defines the freeze-out parameters for a specific hadron. For the pion case, see Fig.~\ref{fig:TimeScalesDiagram}. 

The systematic analysis of the experimental data in the energy range from $1 A$GeV up to $130 A$GeV 
with the statistical model of hadron production assuming universal freeze-out resulted in a set of points in the $T-\mu$ plane. A curve could be fitted which follows the compact parametrization 
\cite{Cleymans:2005xv}
\begin{equation}
T(\mu) = a - b\mu^2-c\mu^4~,
\label{eq:TmuParametrization}
\end{equation}
with $a=0.166\pm0.002$ GeV, $b=0.139\pm0.016~{\rm GeV}^{-1}$
and $c=0.053\pm0.021~{\rm GeV}^{-3}$.
The energy dependence of the freeze-out can be determined by the following parametrization of the
baryochemical potential
\begin{equation}
\mu(\sqrt{s}) = \frac{d}{1+e\sqrt{s}}~,
\label{eq:sdependence}
\end{equation}
with $d=1.308\pm0.028$ GeV and $e=0.273\pm0.008~{\rm GeV}^{-1}$. 


\begin{figure}[h]
	\begin{center}
	\includegraphics[height=.41\textheight]{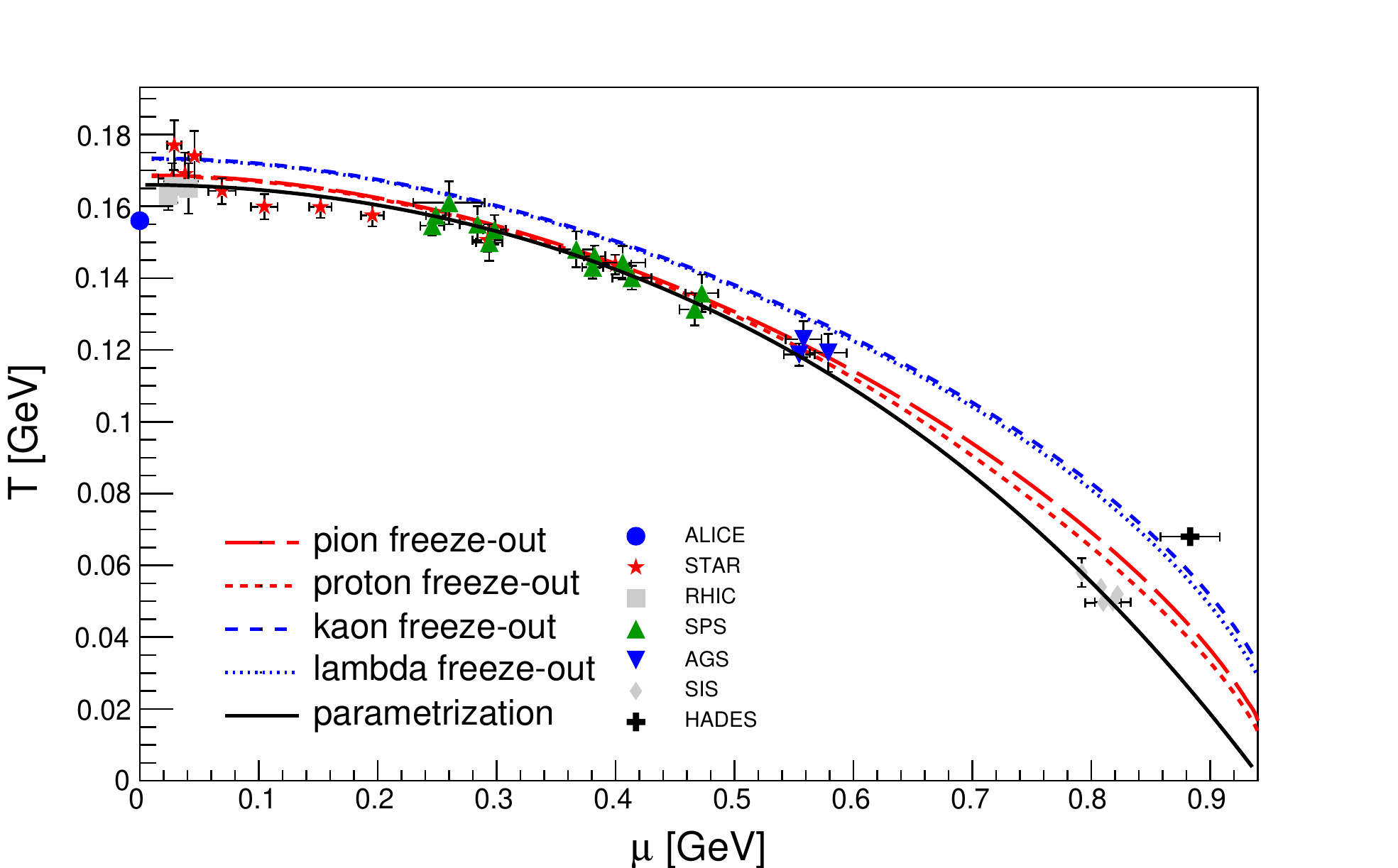}
	\caption{The freeze-out lines for pion, proton, lambda and kaon as determined by our model,
	 and compared to the universal parametrization \cite{Cleymans:2005xv} (solid line).
	 Experimental data at different energies: ALICE \cite{Floris:2014pta},  STAR \cite{Adamczyk:2017iwn,Adams:2005dq},
	 RHIC \cite{BraunMunzinger:2001ip,Cleymans:2004pp}, SPS \cite{Becattini:2005xt,Becattini:2003wp,Bravina:2002wz},
	 AGS \cite{Becattini:2005xt,Becattini:2003wp,Bravina:2002wz}, SIS \cite{Cleymans:1998yb,Averbeck:2000sn}
	 HADES \cite{HADES}.
	}
	\label{fig:freezeout}
	\end{center}
\end{figure}


In Fig.~\ref{fig:freezeout} we show our predictions compared with the parameters extracted from experiments and the curve fitting their systematics.
As a representative set of hadrons we choose pions, nucleons, lambda hyperons and kaons.
For pions and nucleons the freeze-out lines 
practically coincide, and appear in close vicinity  of the experimental points.
In the very same way the freeze-out lines for strange hadrons ($K$ and $\Lambda$)
are close to one another,  attaining freeze-out conditions for slightly larger values of $T$ and $\mu$ as compared to non-strange hadrons. 
The difference in the freeze-out temperatures of $K$ and $\pi$ is about $\Delta T_f\simeq4.7$~MeV at 
$\mu=0$ and stays at this level also for higher $\mu \lsim 0.6$ GeV.
We would like to argue that this proximity lends sufficient support to the assumption of a universal freeze-out line for all hadrons, whereby the quantitative description of strange hadron yields may require correction factors. 
The above universal behaviour of freeze-out for all hadrons is a direct consequence of assumed physical
mechanisms. 
As it is plainly seen in Fig. \ref{fig:freezeout} the parametric line of Eq.~(\ref{eq:TmuParametrization})
is very close to pion and nucleon freeze-out lines determined by our model in the region of high 
temperatures and low densities (the region of high collision energies). 
Significant deviations appear in the low temperature and high density region corresponding to
low energy collisions. 
A notable point was provided by the HADES group \cite{HADES} from the analysis of data from 
Au-Au collisions at $\sqrt{s}=2.4$~GeV energy. 
This point is separated from the line determined by the previous experiments, it has larger values of  $T$ and $\mu$. 
Predictions of our model in the region of high densities and lower temperatures appear closer to the HADES point than to points extracted from the SIS experiment.

\begin{figure}[h]
	\begin{center}
	\includegraphics[height=.41\textheight]{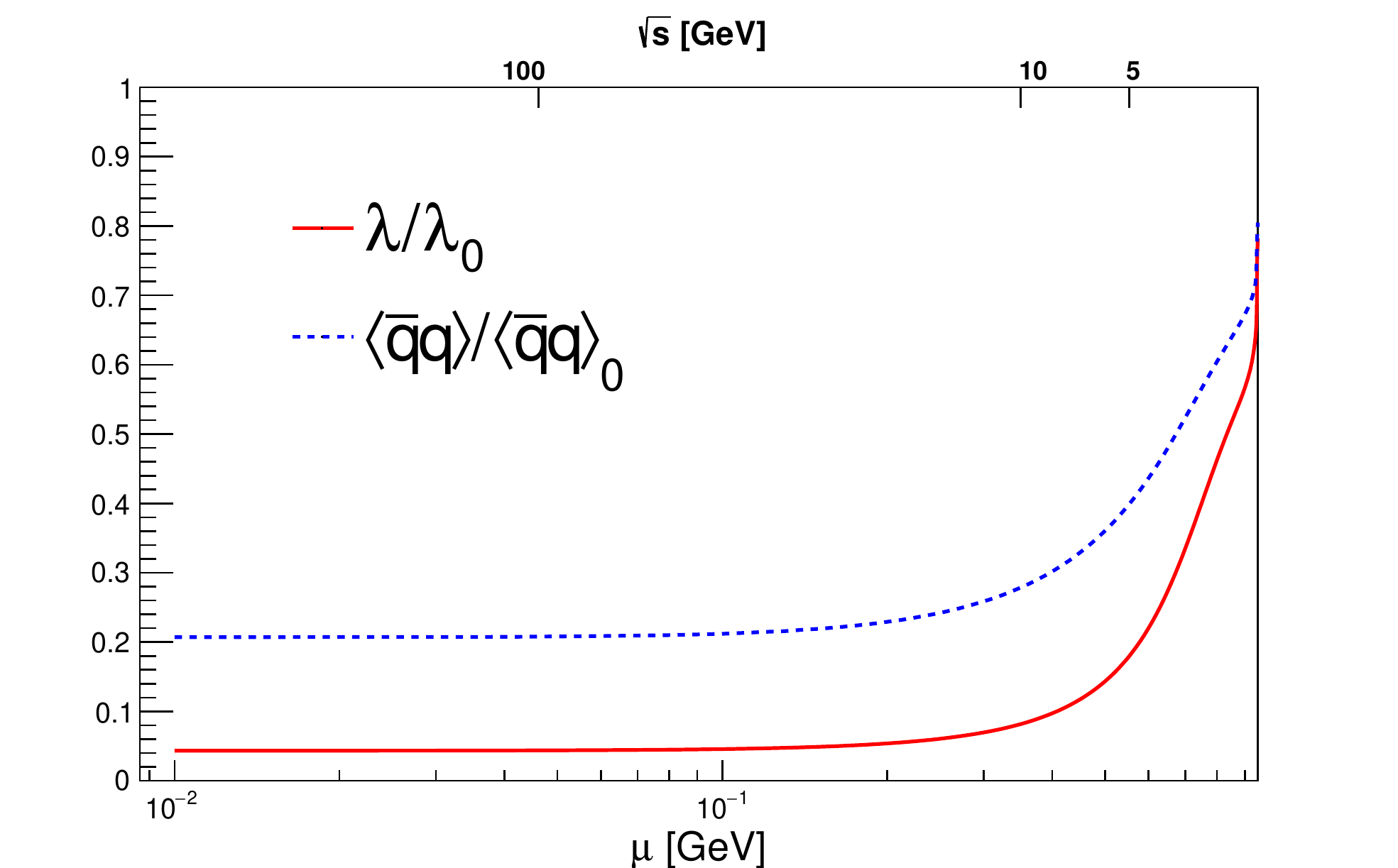}
	\caption{Reduction of the chiral condensate $\langle \bar{q} q\rangle$ and the string tension
	 $\lambda$ relative to their vacuum values ($\langle \bar{q} q\rangle_0$ and $\lambda_0$, resp.) 
	 along the pion freeze-out line defined by Eq.~(\ref{eq:Hexp}). On the upper horizontal
	axis, for orientation, we show the corresponding collision energies according to the fit formula 
	 (\ref{eq:sdependence}).
	}
	\label{fig:Reduction}
	\end{center}
\end{figure}

The freeze-out mechanism adopted in this paper is in  a direct relation with the QCD transition. 
To illustrate this association we plot both quantities along the predicted freeze-out line in 
Fig.~\ref{fig:Reduction}. 
In a wide range of chemical potentials there is a constant, substantial reduction of both order parameters. This strongly suggests that for these chemical potentials and the corresponding values of collision
energies, the freeze-out data should map the QCD transition.
For high densities the chemical kinetics freeze-out condition of Eq.~(\ref{eq:Hexp}) is fulfilled at a much smaller reduction of the QCD order parameters. 
We suggest that this may be due to the different composition of the system which changes its character
from meson dominated at low densities, to baryon dominated at high densities. 
This statement is additionally supported by the behaviour of the baryonic and mesonic contributions to the total entropy density in the hadron resonance gas \cite{Andronic:2009gj} which balance each other at 
$\sqrt{s}\simeq10$~GeV where the transition from baryon dominated to meson dominated matter can be located on the freeze-out line \cite{Andronic:2009gj}.


\begin{figure}[h]
	\begin{center}
	\includegraphics[height=.41\textheight]{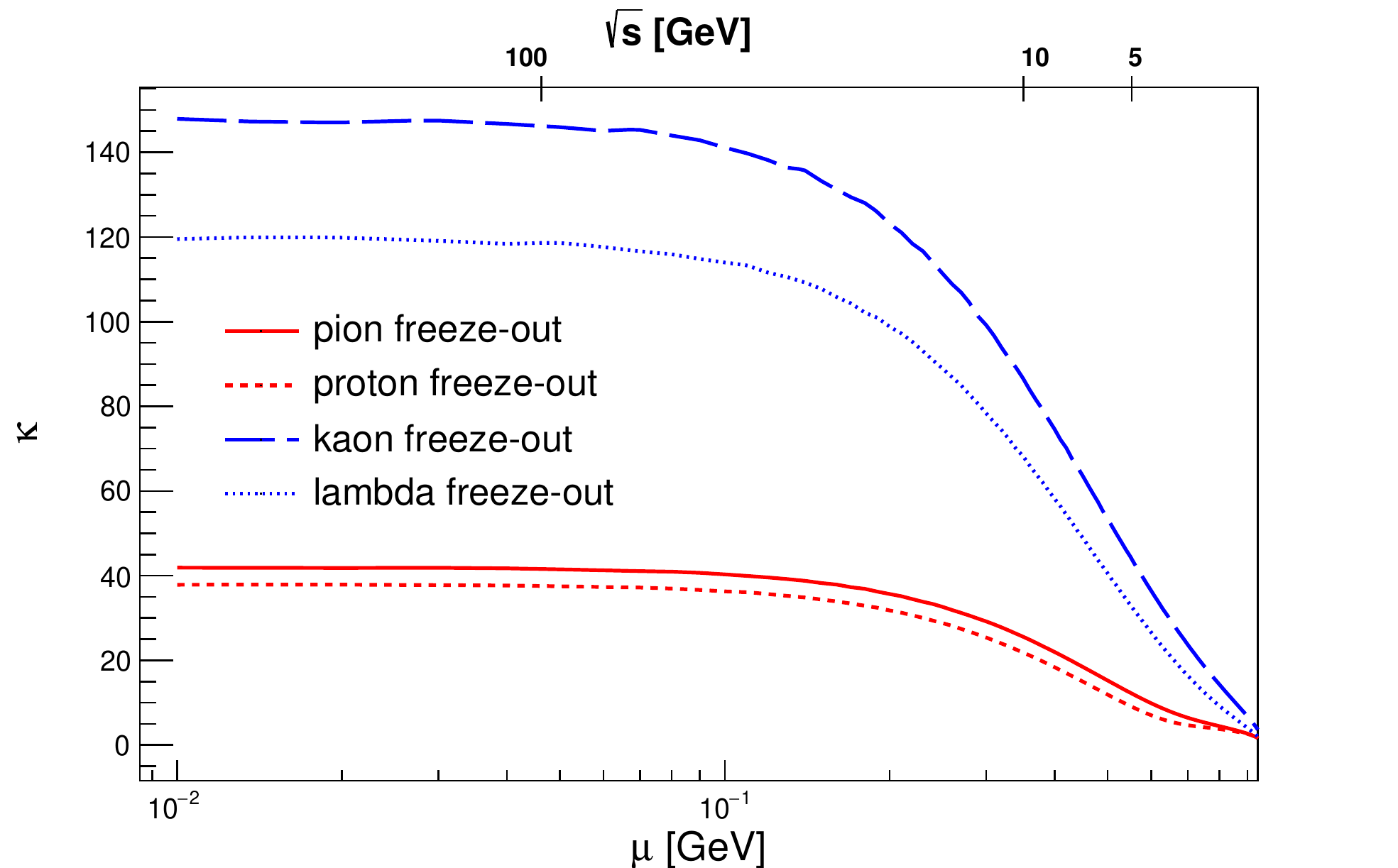}
	\caption{The species dependent exponents $\kappa_i$ for the fit of the temperature dependence of  
	collision rates by a power law $\tau_{{\rm coll},i} \propto T^{\kappa_i}$ defined in 
	Eq.~\ref{eq:KappaEq} along the freeze-out curve for the corresponding species $i=\pi,p,K,\Lambda$.} 
	\label{fig:kappa}
	\end{center}
\end{figure}

Applying a kinetic freeze-out condition like Eq.~(\ref{eq:Hexp}) together with a power law ansatz for the 
temperature dependence of the inelastic collision rate $\tau_{{\rm coll}}\propto T^{\kappa}$,
Heinz and Kestin  \cite{Heinz:2006ur} found that a reasonable description of the centrality independence
of the freeze-out data from the STAR experiment \cite{Adams:2003xp} could be achieved only with 
exponents as high $\kappa\gsim 20$.  
A similarly dramatic temperature dependence was found earlier for the equilibration rate of the 
$Omega$ baryon, $\kappa_\Omega \sim 60$, by Braun-Munzinger et al. \cite{BraunMunzinger:2003zz}.
For a quantitative comparison we extract from the reaction rates defined by our model the 
exponent of a local power-law fit of the temperature dependence by the logarithmic derivative
\begin{equation}
\kappa_i = - \frac{d \ln \tau_{{\rm coll},i}(T,\mu)}{d\ln T}\bigg|_{T_{f,i};\mu_{f,i}}~.
\label{eq:KappaEq}
\end{equation}
In Fig. \ref{fig:kappa} we show the values of the species dependence exponent $\kappa_i$
along the predicted freeze out lines for pions, kaons, nucleons and lambdas. 
In general the values are close to the predictions known from previous studies 
\cite{Heinz:2006ur,BraunMunzinger:2003zz}.
It is plain that for strange hadrons the temperature dependences of the reaction rates are much steeper than for non-strange hadrons, since corresponding values of the exponents are larger. 
This behaviour entails that the collision rates become very short (implying chemical freeze-out) in the close vicinity of the QCD transition.


The conclusion that hadronic reaction rates rapidly drop in a narrow range of temperatures and densities
to provide the chemical freeze-out points to an interpretation that this effect
is intimately related with hadron formation.
Thus, our model supports the physical picture advocated, e.g., in Refs.~\cite{Becattini:1997rv,Heinz:2006ur}, that the chemical freeze-out is a statistical process associated with a phase transition. 
In the present work, we could elucidate the conjecture of Ref.~\cite{Becattini:1997rv} by a calculation based on a model for the in-medium dependence of reaction cross sections and obtained freeze-out parameters in the $T-\mu$ plane from freeze-out for hadrochemical kinetics.


\section{Conclusions}
\label{conclusions}

We have considered a simple model of the chemical freeze-out phenomenon as determined by a kinetic condition applying for each hadronic species.
Two relevant time scales were determined based on well-motivated physical grounds.
The system's expansion rate was estimated from an entropy conserving flow.
The collision time for hadrons was computed from the assumed Povh-H{\"u}fner law
for the scattering cross sections \cite{Povh:1990ad,Hufner:1992cu}, which was generalized for the dense medium by adopting the $T$ and $\mu$ dependence of hadron radii and the string tension.
The size of pions and nucleons was determined using the GMOR relation \cite{GellMann:1968rz} and the NJL prediction \cite{Hippe:1995hu}. 
Those relations were then also used for all the other hadrons. 
The virtue of this choice is that mean squared radii of hadrons depend on the chiral condensate,
allowing for a direct relation of the QCD order parameter and dynamical properties of observable states.
In this picture the physical mechanism driving the chemical freeze-out is a rapid formation of hadrons triggered by {\it chiral symmetry breaking} and {\it colour confinement} appearing at the transition of QCD. We would like to emphasise that as a consequence of our model, the QCD transition in the region of
high baryon densities, as explored in HADES, FAIR or NICA, should be expected in the vicinity of the chemical freeze-out.
Predictions of our model agree with thermal model analysis of the particle multiplicities measured in HIC in a wide energy range.

A natural extension of the model is to build a unified kinetic description of the hadron-hadron reactions, where the freeze-out condition would follow naturally. 
This could, in principle, allow to account for both, the chemical and kinetic freeze-outs. 
This would answer the question whether both effects coincide or are separated in the temperature-density plane.
This approach would naturally include more realistic modelling of hadron-hadron cross section in the 
form of non-relativistic quark exchange contribution \cite{Martins:1994hd,Barnes:1991em} 
relevant for the chemical freeze-out kinetics. \\


{\bf Acknowledgments.}
 We thank Krzysztof Redlich, Olaf Kaczmarek, Gra{\.z}yna Odyniec and Pasi Huovinen  for discussions.
 The research was supported
 by the NCN grant No. UMO-2014/15/B/ST2/03752 (DB, JJ)
and by the MEPhI Academic Excellence Project under contract No. 02.a03.21.0005 (DB). 


\end{document}